\documentclass{elsart}

\usepackage{graphicx}
\usepackage{dcolumn}
\usepackage{bm}
\usepackage{amssymb}

\begin{document}

\begin{frontmatter}


\title{ Surface instabilities in granular matter and ion-sputtered surfaces }

\author{ T. Aste$^{1,2}$}
\author{U. Valbusa$^2$ }
\address{$^1$ Applied Mathematics, RSPHYSSE, ANU, 0200 Canberra ACT, Australia }
\address{$^2$ INFM-Dipartimento di Fisica, Universit\`a di Genova, via Dodecaneso 33, 16146 Genova, Italy}


%

%
\begin{abstract}

We apply a theoretical approach, originally introduced to describe aeolian ripples formation in sandy deserts, to the study of surface instability in ion sputtered surfaces.
The two phenomena are distinct by several orders of magnitudes and by several physical mechanisms, but they obey to similar geometrical constraints and therefore they can be described by means of the same approach. 
This opens a novel conceptual framework for the study of the dynamical surface roughening and ripple formation on crystal and amorphous surfaces during ion sputtering.
\end{abstract}

\end{frontmatter}


\section{Introduction}

In the literature, many studies have been devoted to understanding the mechanism of aeolian ripple formation \cite{Bagnold,Andersen02,Andersen01,Kurtze00,Hoyle99,Prigozhin99,Werner93}.
In particular, a hydrodynamic(al) model, based on a continuum dynamical description with two species of grains (immobile and rolling grains), was proposed with success by Bouchaud et al. \cite{Bouch95,Terz98,Val99,Csa00}. 
The main ingredient of such a model is a bilinear differential equation, for the population of the two species of grains, which shows the instability of a flat bed against ripple formation.
In this paper we show that the same reasoning which has been used to describe the sand ripples formation in deserts can be conveniently applied to the studies of dynamical surface roughening \cite{Ekl91,Krim93,Chra94,Valbusa02,Buat00,Rusp98,Rusp98a,Rusp97} leading to an accurate description of the morphogenesis and evolution of ripples on crystal and amorphous surfaces during ion sputtering.
To this end we substantially extended the original approach by introducing new terms describing the particle-mobility on the surface.
In ion-sputtered surface growth, these new terms play a central role as control parameters in the dynamics of ripple formation.
The present approach contains the Bradley-Harper approach \cite{Bra88,Par99} (based on a Kardar-Parisi-Zhang type equation \cite{KPZ}) and it represents an alternative, independent and original way to study the problem of surface instabilities.

\begin{figure}
\includegraphics[width=12cm]{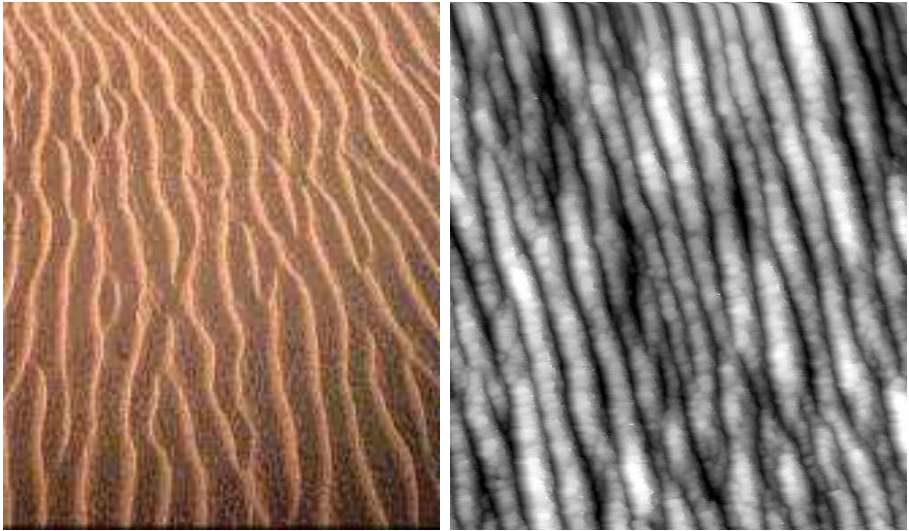}
\caption{\label{f.0} 
Two examples of ripple formation in two completely distinct physical systems: Ripples on sand (Gobi desert) and Ripples on surfaces (Ag under ion sputtering).
The line below the two figures corresponds approximately respectively to 1 {\it m} and 50 {\it nm} }
\end{figure}

\section{Excavation, Adsorption and Mobility}

When the surface of a solid is taken under ion sputtering some atoms in the proximity of the surface receive energy from the sputtered ions and pass from a bounded - \emph{`immobile'} - solid state to a \emph{`mobile'} unbounded state.
The opposite mechanism is also allowed: some \emph{mobile} atoms can gain in energy by becoming \emph{immobile} and bounding in a given position in the solid.
A certain fraction of atoms might also be dispersed into the atmosphere. 
Let us call $h({\mathbf r },t)$ the height of surface profile made of immobile -bounded- atoms and call $R({\mathbf r },t)$ the height of mobile -melted- atoms.
In analogy with the theory developed to explain the dynamical evolutions of dunes in deserts \cite{Bouch95,Terz98,Val99,Csa00}, we describe the mechanisms of excavation, exchange between mobile and immobile atoms and surface displacement of mobile atoms in term of the following differential equation:
\begin{eqnarray} \label{E1}
\frac{\partial h}{\partial t} 
&=&  - \Gamma(R,h)_{ex} + \Gamma(R,h)_{ad} 
\nonumber \\
\frac{\partial R}{\partial t} 
&=& - {\mathbf \nabla} {\mathbf J}(R,h) + (1-\phi) \Gamma(R,h)_{ex} -\Gamma(R,h)_{ad} \;\;\;.
\end{eqnarray}
Where $\Gamma(R,h)_{ex}$ is the rate of atoms that are excavated under the action of the sputtering, and $(1-\phi)$ is the part of them that pass from immobile to 
mobile, whereas $\phi$ is the fraction that is dispersed into the atmosphere. 

Let us now write in details the various terms contained in Eq.\ref{E1}.

First we consider the the excavation effect which must depend on the number and velocity of the sputtered ions (i.e. its flux), but also the local shape and orientation of the surface might play an important role.
Indeed, the energy transmitted by the impacting ions concentrate more in regions of the surface with positive curvature. 
Moreover, part of the surface facing the flux are likely to experience a different erosion respect to others which are less exposed to the flux.
Crystalline orientation and anisotropies might be also taken into account.
We can write:
\begin{equation}\label{Gex}
\Gamma(R,h)_{ex} = 
\eta 
( 
1 
+ {\mathbf  a }    {\mathbf \nabla} h 
+ { b }    {\mathbf \nabla}^2 h  
) \;\;\; ;
\end{equation}
here $\eta$ is the sputtering flux, whereas ${\mathbf a}$ and ${ b }$ are respectively associated with the flux-direction-dependent and with the curvature-dependent sputtering erosion.

We now consider the adsorption process. First we note that the rate of adsorption of mobile atoms into immobile solid positions must be dependent on the quantity of mobile atoms in a given spatial position.
Similarly to the excavation process, the adsorption is also dependent on the local curvature and orientation. 
We can write:
\begin{equation}\label{Gad}
\Gamma(R,h)_{ad} =  
R 
( 
\gamma 
+ {\mathbf  c }    {\mathbf \nabla} h 
+ { d }    {\mathbf \nabla}^2 h  
) \;\;\;,
\end{equation}
where the parameter $\gamma$  is the recombination rate and ${\mathbf c}$ and ${ d }$  are associated to the different probabilities of recombination in relation with the local orientation and shape of the surface. 

Note that Eqs. \ref{Gex}, \ref{Gad} contain the same terms as the ones proposed in the literature for the formation of aeolian dunes in the so-called hydrodynamical model \cite{Bouch95,Terz98,Val99,Csa00,Mak00}.
Indeed, in deserts, sand grains are lifted from the sand-bed and readsorbed into it with a probability which is dependent on the local shape and orientation of the dunes.
Eqs. \ref{Gex}, \ref{Gad} \emph{represent the simplest analytical expressions which formally take into account these shape and orientation dependences}.
In the search for simple explanations, such equations are therefore rather universal.

In this paper we do not intend to enter into the details of the physics of surface erosion and adsorption. 
Further discussion on can be found in the literature \cite{Par99,KPZ,Muth84,Rost95,Kur84,Meakin86,Rever98,Sanc01,Zhur03}.

Mobile atoms will move on the surface, and the quantity ${\mathbf J}(R,h) $ in Eq.\ref{E1} is the `current' of these atoms. 
In surface growth, there are two main mechanisms that are commonly indicated as the responsible for the surface mobility of atoms \cite{Bar95}.
The first is a current, driven by the variations of the local chemical potential, which tends to smoothen the surface asperity moving atoms from hills to valleys. The second is the current induced by the Erlich-Schwoebel barrier which -on the contrary- moves atoms uphill.
In addiction to these main mechanisms we might also have to take into account a drift velocity and a random thermal diffusion, obtaining:
\begin{equation} \label{J1}
{\mathbf J}(R,h)  = 
{\mathcal K} R {\mathbf \nabla}({\mathbf \nabla}^2 h)
+ s R  \frac{ {\mathbf \nabla} h }{1+ ( \alpha_d {\mathbf \nabla} h)^2}
+
{\mathbf v } R 
- {\mathcal D } {\mathbf \nabla } R 
  \;\;\;.
\end{equation}
In this equation, the first term describes a deterministic diffusion driven by the variations of the chemical potential which depends on the local shape of the surface; the second term is associated with the uphill current due to the Erlich-Schwoebel barrier and $\alpha_d$ is a constant associated with the characteristic length.
The quantity ${\mathbf v }$ is a drift velocity of the mobile atoms on the surface, whereas ${\mathcal D }$ is the dispersion constant associated with the random thermal motion.

Note that Eq.\ref{J1} \emph{is substantially different from the one proposed in the literature to describe ripples in granular media} \cite{Bouch95,Terz98,Val99,Csa92,Csa93}. 
Here the current is supposed to be dependent on the local shape and orientation of the surface (the $ h({\mathbf r },t)$ profile). 
The equations describing sand deserts can be retrieved from Eq.\ref{J1} by imposing ${\mathcal K}=0$ and $s=0$, but -on the contrary- in surface growth these two parameters are the leading terms of the equation and play the role of control parameters in the dynamics of ripple formation.
Nonetheless, these terms describe a rather simple dependence of the dynamic of particles on a surface on the geometrical shape of the surface itself.
Again, in our seek for universality, we expect that similar terms can be profitably introduced in the context of aeolian sand ripples in order to describe specific phenomena (associated, for instance, with packing properties \cite{Aste00} or granular flow \cite{Komatsu01}) which relate the current of grains with the dune-shapes.

It should be noted that the factors ${\mathbf a}$, ${\mathbf c}$ and ${\mathbf v}$ in Eqs.\ref{Gex},\ref{Gad} and \ref{J1} are \emph{vectors} (i.e. they have -in general- different components in the two horizontal directions).
Indeed, crystal surfaces are in general anisotropic and therefore one must take into account the dependence of the parameters on the relative orientation of the crystal-surface and the sputtering direction.
However in this paper we focus on the 1-dimensional case only.
Preliminary results show that this approach can be indeed straightforwardly extended to the two-dimensional case.

\begin{figure}
\includegraphics[width=12cm]{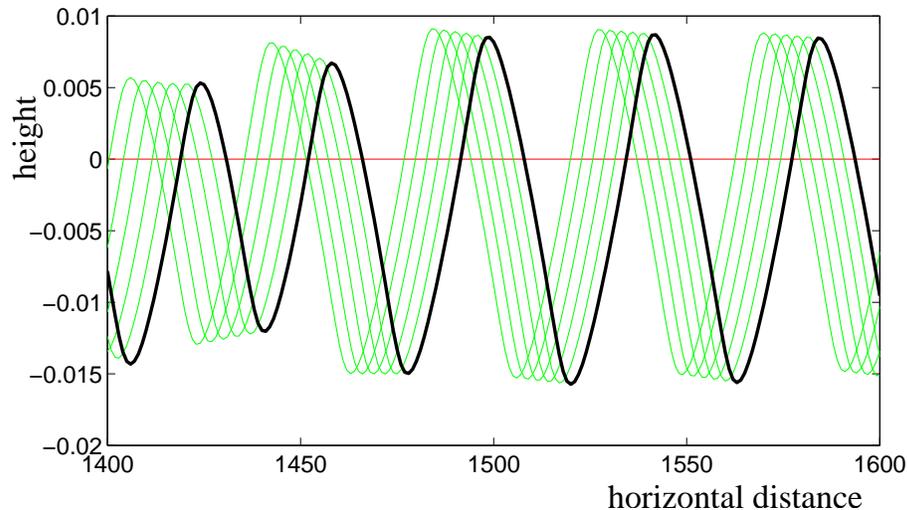}
\caption{\label{f.1} 
Numerical solutions of Eq.\ref{E1} at various times indicate that under the action of ion sputtering the surface develops an instability which leads to the formation of ripples with a well defined characteristic wavelength.
In the figure the black-thick line is the final surface-profile, whereas the thinner-gray (-green, online version only) lines are some profiles at previous times. (See Appendix \ref{A.sim} for details).}
\end{figure}

\section{Dispersion Relation}

A trivial solution of Eq.\ref{E1} can be written for a completely flat surface: $h({\mathbf r },t)=h_0(t)$ and $R({\mathbf r },t)=R_0$.
In this case, we obtain $R_0 = (1-\phi) \eta/ \gamma$ and $h_0(t)=-\phi \eta t + const.$.
This describes a surface that rests flat and it is eroded with a speed equal to $\phi \eta$.
But this behavior is only hypothetical since -in general- the dynamics of the surface-profile presents instabilities against spontaneous roughening and therefore its evolution is more complex.
For instance, a numerical solution of Eq.\ref{E1}, is shown in Fig.\ref{f.1} (for the 1-dimensional case).
We observe that, in a certain range of the parameters, the surface is unstable and periodic ripples are formed spontaneously.


In order to infer indications about the amplification or the smoothing of small perturbations and to deduce an analytical expression for the ripples wave-length  at their beginning, we perform a stability analysis on Eq.\ref{E1}.
For this purpose we assume that the surface-profile is made by the combination of a flat term plus a rough part: 
\begin{eqnarray}\label{R1h1}
R({\mathbf r },t) &=& R_0 + R_1({\mathbf r },t) \nonumber \\
h({\mathbf r },t) &=& h_0(t) + h_1({\mathbf r },t) \;\;\;,
\end{eqnarray}
with $ R_1({\mathbf r },t) = \hat R_1 \exp(i \omega t + i {\mathbf k } {\mathbf r })$ and $ h_1({\mathbf r },t) = \hat h_1 \exp(i \omega t + i {\mathbf k } {\mathbf r })$. 
We substitute these quantities into Eq.\ref{E1} and linearize the equation by taking only the first order in $R_1$ and $h_1$.
A Fourier analysis (see Appendix \ref{A1}) shows that such a linearized equation admits solutions when the frequencies $\omega$ and the wave vectors ${\mathbf k}$ satisfy:
\begin{eqnarray} \label{Deter}
& &\left[ 
i \omega + 
\gamma 
+ i {\mathbf k } {\mathbf v } 
+  k^2 {\mathcal D } 
\right] 
\left\{
i \omega 
+  i {\mathbf k } \left[ {\mathbf v }_1 - (1-\phi) {\mathbf v }_2 \right]
- k^2 \left[ {\mathcal D }_1 -(1-\phi) {\mathcal D }_2 \right]
\right\}-
\nonumber \\
& &
\gamma
(1-\phi)
\!
\left[
\!
i {\mathbf k } ({\mathbf v }_1 
\! - \!{\mathbf v }_2) 
\! - \! k^2  \left({\mathcal D }_1 
\! - \! {\mathcal D }_2 - s_1 \right)
\! - \! k^4 {\mathcal K }_1
\right] = 0
\;  ;
\nonumber \\
\end{eqnarray}
where, to simplify the equations, we have introduced the following notation: 
\begin{center}
\begin{tabular}{lll}
${\mathbf v}_1 = \eta {\mathbf a}$ 
\hspace{1.cm} &  ${\mathbf v}_2 = \eta {\mathbf c}/{\gamma}$ \\
${\mathcal D }_1 = \eta b$
& ${\mathcal D }_2 = \eta d/{\gamma}$ \\
$s_1 = \eta s/{\gamma}$
& ${\mathcal K }_1 = \eta {\mathcal K }/\gamma$ 
\end{tabular}
\end{center}
Equation \ref{Deter} establishes a \emph{dispersion relation} $\omega({\mathbf k})$ that is a complex function with two branches corresponding to the solutions of the quadratic Equation \ref{Deter}.
\begin{figure}
\includegraphics[width=12cm]{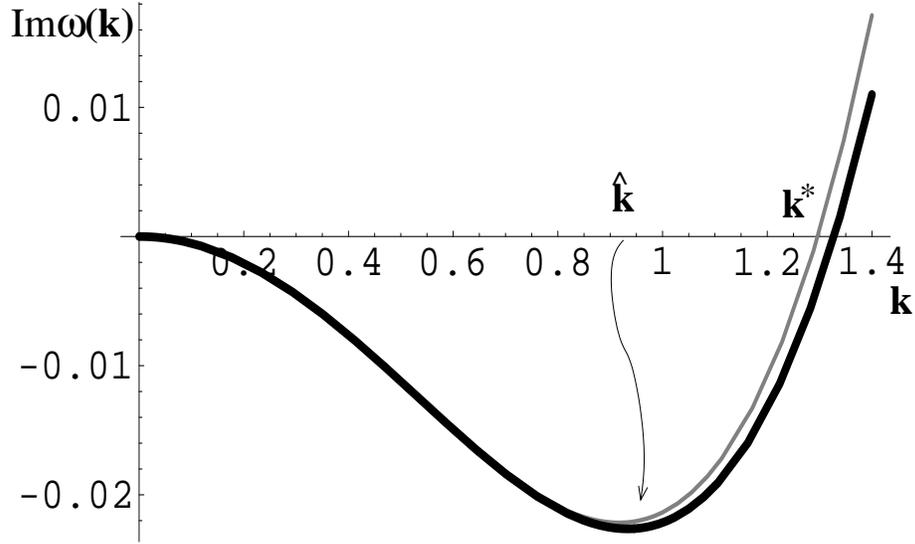}
\caption{\label{f.2} 
The imaginary part of the dispersion relation $Im(\omega)$ can assume negative values which are associated with the surface instability (arbitrary units) \cite{typical}.
The amplitude of modes with wavelengths $\lambda > 2\pi/k^*$ will grow exponentially fast.
The thick line is the imaginary part of the analytical solution of Eq.\ref{Deter}, whereas the thinny-gray line is the approximated expression (at the fourth order in $k$) obtained for small ion flux ($\eta$ small). 
This figure is representative of a rather general behavior within a large range of parameters. (For this figure we used: $\gamma = 8$, $\phi = 0.5$, $v = 0.4$, $\mathcal D = 0.2$, ${\mathcal K}_1 = 0.06$, $s_1 = 0.09$, $v_1=0.01$, $v_2=0.001$, ${\mathcal D}_1 = 0.015$ and ${\mathcal D}_2 = 0.023$.)  
}
\end{figure}

\section{Surface Instabilities}

The kinetic growth of the surface instability is related to the imaginary part of $\omega({\mathbf k})$.
Indeed, $Im(\omega({\mathbf k}))$ corresponds to modes with amplitudes that change exponentially fast with time, and negative values correspond to unstable modes that increase with the time. 
We can therefore study $Im(\omega)$ from the solution of Eq.\ref{Deter} and search for the range of $k$ in which $Im(\omega)$ is negative.
The most unstable mode is the one that grows faster and it corresponds to the value of ${\mathbf k}$ at which $Im(\omega)$ reaches its most negative value (see Fig.\ref{f.2}).

The solution of Eq.\ref{Deter} for $Im(\omega)$, is 
\begin{equation}\label{solut}
2 Im(\omega)_\pm 
 = 
\gamma +
 \Big[{\mathcal D } - {\mathcal D }_1 +(1-\phi) {\mathcal D }_2 \Big]k^2
 \pm 
\sqrt{\frac{\Delta_1 + (\Delta_1^2 + 4 \Delta_2^2)^{1/2}}{2}}
\end{equation}
where we have
\begin{eqnarray}\label{delta1}
\Delta_1 &=& 
\gamma^2 - \Big\{ \Big[ {\mathbf v} - {\mathbf v}_1 +(1-\phi) {\mathbf v}_2\Big] {\mathbf k} \Big\}^2
\nonumber \\
&+&
2 \gamma \Big[ {\mathcal D } - (1-2\phi){\mathcal D }_1 + (1-\phi) {\mathcal D }_2 + 2 (1-\phi)s_1\Big] k^2
\nonumber \\
&+& 
\Big\{
	\Big[({\mathcal D } + {\mathcal D }_1 - (1-\phi){\mathcal D }_2\Big]^2 
	- 4 \gamma (1-\phi) {\mathcal K }_1
\Big\} k^4
\nonumber \\
\end{eqnarray}
and 
\begin{eqnarray}\label{delta12}
\Delta_2 &=& 
\gamma \Big[{\mathbf v}+ (1-2\phi){\mathbf v}_1-(1-\phi){\mathbf v}_2\Big] {\mathbf k }
\nonumber \\
&+&
\Big[{\mathcal D } + {\mathcal D }_1- (1-\phi){\mathcal D }_2\Big] 
\Big[{\mathbf v}-{\mathbf v}_1+(1-\phi){\mathbf v}_2 \Big]{\mathbf k }^3 \;\;.
\nonumber \\
\end{eqnarray}

Let us first observe that in absence of sputtering (i.e. when $\eta = 0$ and therefore, ${\mathbf v }_1=0$, ${\mathbf v }_2=0$, 
${\mathcal D }_1=0$, ${\mathcal D }_2=0$,
$s_1=0$, ${\mathcal K }_1=0$) the solutions of Eq.\ref{Deter} are $\omega({\mathbf k}) =0$ and $\omega({\mathbf k}) = - {\mathbf k } {\mathbf v } + i (\gamma + k^2 {\mathcal D })$.
In this case, the imaginary part of $\omega({\mathbf k})$ is non-negative, therefore we -correctly- expect no spontaneous corrugation of the surface.
On the contrary, when the sputtering is active ($\eta \not= 0$), the imaginary part of $\omega({\mathbf k})$ can assume negative values as shown in Fig.\ref{f.2} where a plot of $Im(\omega)_-$ is reported.
As one can see, typically the branch $Im(\omega)_-$ takes negative values for $k$ between 0 and a critical value $k^*$ at which it passes the zero \cite{typical}.
The critical point $k^*$, fixes the minimal unstable wavelength.
We therefore expect to find unstable solutions associated with the formation and evolution of ripples with wavelengths $\lambda \ge \lambda^* = 2\pi/k^*$.

\section{Ripple wavelength}

Several analytical solutions of Eq.\ref{Deter} can be found in special cases which are discussed in Appendix \ref{A2}.
But the study of the surface instabilities can be highly simplified if we consider the first order effects when the sputtering flux $\eta$ is small.


In the case of small sputtering fluxes, the branch of $Im(\omega({\mathbf k}))$, with negative values can be approximated to:
\begin{equation}\label{dev}
Im(\omega)_-   
\simeq  
\frac{
P_1 k^6 + P_2 k^4 + P_3({\mathbf k})k^2 + P_4 k^2 + P_5({\mathbf k}) }{
{\mathcal D }^2 k^4 
+ 2 \gamma {\mathcal D } k^2 
+ ({\mathbf v  \mathbf k})^2
+\gamma^2  
}
\end{equation}
with
\begin{eqnarray}
&&P_1 = {\mathcal D } [ (1-\phi)\gamma {\mathcal K }_1 - {\mathcal D }({\mathcal D }_1 -(1-\phi) {\mathcal D }_2 )]
\nonumber \\
&&P_2 = (1-\phi) \gamma  [{\mathcal D }({\mathcal D }_2-s_1) + \gamma {\mathcal K }_1] - (1+\phi)\gamma {\mathcal D }{\mathcal D }_1  \nonumber \\
&&P_3({\mathbf k}) = - [{\mathcal D }_1 + (1-\phi) {\mathcal D}_2 ] ({\mathbf v}{\mathbf k})^2
\nonumber \\
&& P_4 = \gamma^2 [ (1-\phi) s_1 - \phi {\mathcal D }_1 ] 
\nonumber \\
&& P_5({\mathbf k}) = (1-\phi) \gamma ({\mathbf v}{\mathbf k}) [({\mathbf v}_1-{\mathbf v}_2){\mathbf k}] 
\end{eqnarray}

When $k$ is sufficiently small ( $k \ll \gamma/\eta$ ), we can develop  Eq.\ref{dev} at the 4$^{th}$ order obtaining:
\begin{equation}\label{4ord}
Im(\omega)_-  \simeq  A k^4 - B k^2 \;\;,
\end{equation} 
with
\begin{eqnarray}\label{AB}
A &=& (1-\phi) \Big[
{\mathcal K }_1 
+ (s_1 + {\mathcal D }_2 -{\mathcal D }_1) \frac{ \gamma {\mathcal D } + v^2}{\gamma^2}+
v (v_1-v_2) \frac{ 2 \gamma {\mathcal D } + v^2}{\gamma^3}
\Big]
\nonumber \\
B &=& \phi {\mathcal D }_1 + (1-\phi) \Big[ s_1  + \frac{ v (v_1-v_2)}{\gamma} \Big]\;\;.
\end{eqnarray}
Here $v$, $v_1$  and $v_2$ are respectivelly the components of ${\mathbf v}$, ${\mathbf v}_1$  and ${\mathbf v}_2$ in the direction parallel to ${\mathbf k}$).
(In Fig.\ref{f.2} a comparison between this approximate solution and the exact one is given.)


The expected wavelength of the ripples is associated with the fastest growing mode, which corresponds to the value of ${\mathbf k}$ at which $Im(\omega)_-$ reaches its most negative point.
Here the minimum of $Im(\omega)$ is at 
\begin{equation}\label{khat}
\hat k = \sqrt{\frac{B}{2A}}\;\;\;.
\end{equation}
Therefore, at the beginning, the roughness will grow exponentially fast as $W \sim \exp(B^2 t/(4A))$ with associated ripple-wavelength at: \begin{equation}\label{wave}
\hat \lambda \sim 2 \pi \sqrt{\frac{2 A }{B }} \;\;\;\;.
\end{equation}


Let us now study some special cases. We first observe that, when ${\mathcal K }_1 $, $s_1$ and $\phi $ are all equal to zero, the ripple wavelength, given by Eq.\ref{wave}, coincides with the one found for sand dunes in deserts (see for instance \cite{Val99}). 
In our notation the `reptation length' is $l_0 = v/\gamma$, the `cut-off length' is $l_c = ({\mathcal D }_2-{\mathcal D }_1)/v$, whereas $v_1-v_2$ is the collective drift velocity of the dunes.
The approximations usually applied in this context \cite{Terz98,Val99}, imply: $l_c \gg \sqrt{{\mathcal D }/\gamma}$, and $\gamma l_c \gg v_1-v_2$.
Giving, from Eq.\ref{wave}
\begin{equation}
\hat \lambda \sim 2 \pi \sqrt{ \frac{2 v l_0 l_c }{v_1-v_2} } \;\;\;\;.
\end{equation}

Let us now consider the dynamical evolution of a surface under ion sputtering and in particular the case when the effect of the Erlich-Schwoebel barrier is not present (as for semiconductors and glasses). 
In this case, $s=0, s_1=0$ and we also expect that the drift velocity $v$ and the dispersion constant ${\mathcal D }$ are equal to zero or infinitesimally small.
Indeed, here the current of mobile atoms on the surface is mainly induced by the differences in the chemical potential. 
Under these assumptions, from Eq.\ref{wave}, the wavelength of the most unstable ripple is:
\begin{equation}\label{lk}
\hat \lambda \sim 2 \pi \sqrt{ \frac{2 {\mathcal K }}{\nu}} \;\;\;\;.
\end{equation}
where we called $\nu = \gamma b \phi/(1-\phi) $, a quantity which plays the role of an effective surface tension.
Note that Equation \ref{lk} is the same result as from the Bradley and Harper theory \cite{Bra88,Par99,Bar95,Cuer95,Gill01}.

When the Erlich-Schwoebel barriers are active ($s,s_1 \not= 0$), effects can be observed on the ripple-wavelength at their beginning, which becomes:
\begin{equation} \label{lks}
\hat \lambda \sim 2 \pi \sqrt{\frac{2 {\mathcal K }}{\nu + s }} \;\;\;\;.
\end{equation}

\section{Conclusions}

We have shown that the same theoretical approach introduced to describe the formation of aeolian sand ripples can be conveniently applied to the study of the formation of periodic structures on surfaces under ion sputterning.
Although the two phenomena are rather different, they can be described within the same conceptual framework by using rather general ideas that relate mobility, excavation and adsorption rates with the surface shape and orientation.
The main purpose of this paper is to point out a relevant example of universality: two processes which have completely different scales and underlying physical mechanisms present a dynamical evolution which \emph{obeys to the same geometrical constraints} and thus can be described by using the same phenomenological model.
On the other hand, we must observe that the class of solutions of Eq.\ref{E1} is rich and complex - even in the linear approximation.
Exhaustive, systematic studies of the classes of solutions of this equation and their dependence on the set of parameters will be the subject of future studies and publications.


\appendix

\section{Fourier transform of the linearized equation} \label{A1}

By substituting Eqs.\ref{R1h1}, \ref{J1}, \ref{Gad} and \ref{Gex} into Eq.\ref{E1} and by neglecting the second order terms (in $R_1$ and $h_1$), we obtain the following linearized equation:
\begin{eqnarray} \label{EqStab}
\frac{\partial h_1}{\partial t} 
&=&  
   \gamma R_1 
- \left[ {\mathbf v}_1 - (1-\phi){\mathbf v}_2 \right] {\mathbf \nabla} h_1 
- 
\left[ {\mathcal D }_1 - (1-\phi){\mathcal D }_2 \right] {\mathbf \nabla}^2 h_1 
\nonumber \\
\frac{\partial R_1}{\partial t} 
&=& 
- \gamma R_1
- {\mathbf v }  {\mathbf \nabla } R_1 
+ {\mathcal D } {\mathbf \nabla }^2 R_1 +
\nonumber \\
&& (1-\phi)\Big[ ({\mathbf v}_1-{\mathbf v}_2)   {\mathbf \nabla}  h_1 +   
 ({\mathcal D }_1 -{\mathcal D }_2 - s_1    ) {\mathbf \nabla}^2 h_1
- {\mathcal K }_1 {\mathbf \nabla}^4 h_1 \Big]
\;\;\;\;.
\end{eqnarray}

A Fourier analysis of Eq.\ref{EqStab} leads to
\begin{eqnarray} \label{EqFou}
& &
\gamma  \hat R_1 
-\Big\{
i \omega 
+  
i {\mathbf k } \left[ {\mathbf v }_1 - (1-\phi) {\mathbf v }_2 \right] -
k^2 \left[ {\mathcal D }_1 -(1-\phi) {\mathcal D }_2 \right]
\Big\} \hat h_1 = 0
\nonumber \\
& &
\left[ 
i \omega + 
\gamma 
+ i {\mathbf k } {\mathbf v } 
+  k^2 {\mathcal D } 
\right]  \hat R_1 
-
(1-\phi)
\Big[
  i {\mathbf k } ({\mathbf v }_1 -{\mathbf v }_2) 
-
\nonumber \\
& &
k^2  \left({\mathcal D }_1 - {\mathcal D }_2 - s_1 \right)
- k^4 {\mathcal K }_1
\Big] \hat h_1 = 0
\;\;\;,
\end{eqnarray}
with $\hat R_1$ and $\hat h_1  $ the Fourier components of $R_1$ and $h_1$ respectively.
This equation is a simple linear equation in two variables. 
It admits a non-trivial solution when the determinant of the coefficients is equal to zero.
This leads to Eq.\ref{Deter}.

\section{Exact solutions} \label{A2}

Analytical expressions for the value of $\mathbf k$ at which $Im(\omega)=0$ (${\mathbf k}^*$) can be calculated from Eq.\ref{solut} in some special cases. 

In particular, when $\phi=0$, $s_1=0$, ${\mathcal K }_1=0$ and ${\mathcal D }=0$, we obtain
\begin{equation}\label{a0}
k^*
=
\sqrt{
\frac{
 \gamma ({v }_1 -{ v }_2)   }{ 
(v - { v }_1 + { v }_2)
 ({\mathcal D }_2 - {\mathcal D }_1) 
} 
}  
\;\;\;,
\end{equation}
where $v$, $v_1$ and $v_2$ are the components of ${\mathbf v}$, ${\mathbf v}_1$ and ${\mathbf v}_2$ in the direction of ${\mathbf k }^*$.

On the other hand when, ${\mathcal K }_1$, $s_1$, ${\mathcal D }_1$ and ${\mathcal D }_2=0$, we find 
\begin{equation}\label{a3}
k^*
=
\sqrt{
\frac{
\gamma ({v } - \phi { v }_1)
}
{
\mathcal D ({v }_1 - { v } -(1-\phi) { v }_2) 
}} \;\;\;.
\end{equation}

The effect of the deterministic diffusion induced by the chemical potential can be studied from the solution  \begin{equation}\label{a1}
k^*
=
\sqrt{
\frac{\phi \gamma {\mathcal D }_1}{ 
(1-\phi) \gamma {\mathcal K }_1  - {\mathcal D }({\mathcal D }_1 - (1-\phi) {\mathcal D }_2) } }  \;\;\;,
\end{equation}
which holds when ${\mathbf v }=0$, ${\mathbf v }_1=0$, ${\mathbf v }_2=0$, $s_1=0$ and ${\mathcal D }-{\mathcal D }_1 + (1-\phi) {\mathcal D }_2 >0$.

Whereas, when ${\mathbf v }_1$, ${\mathbf v }_2$, ${\mathcal D }_1$ and ${\mathcal D }_2=0$, we find 
\begin{equation}\label{a2}
k^*
=
\sqrt{
\frac{ s_1 }{ 
{\mathcal K }_1  } }  \;\;\;,
\end{equation}
\noindent
which implies that the uphill current due to the Erlich-Schwoebel barrier can generate instability even when the shape-dependent erosion and recombination terms are inactive.

\section{Numerical Simulations} \label{A.sim}

The numerical solutions of Eq.\ref{E1} presented in this paper and in particular the ones shown in Figs.\ref{f.1} have been performed as follows.
We considered a one-dimensional flat substrate ($h(x,0)=h0$) of length $L$, with periodic boundary conditions. 
An infinitesimal quantity of mobile atoms were added randomly to the substrate (with $0 < R(x,0) < L/N 10^{-13}$).
The profile-evolution was then computed by using Eq.\ref{E1} with the derivative substituted with finite differences. 
To this purpose, the substrate has been divided into $N$ discrete points.
The -adimensional- time is the number of numerical steps.
The height is in unit of $L/N$ and the roughness is defined as $w(t,L)=\left< [h(x',t)- \left< h(x,t) \right>_x]^2 \right>^{1/2}_{x'}$ (see, for instance, \cite{Csa92}).

Several computations with a number of points equal to $N=1000$, 2000, 3000 and 10000 (the one presented here have $N=3000$) have been performed to verify the effect of boundary and discretization.
Moreover, simulations with no periodic boundary conditions and with the sputtering term (Eq.\ref{Gex}) applied only to a central mask, have also been performed obtaining very similar results.
The robustness of the present approach has been verified varying the parameters, the time steps, the initial roughness of the substrate, etc.
Comparable results have been always found but, we must stress that, under some conditions, numerical instabilities (in particular surface-deformations with $\lambda \sim L/N$) can be trigged on depending on the protocol utilized.

The simulation result shown in Fig.\ref{f.1} uses: $v = 0.05$, ${\mathcal D} = 0.1$, ${\mathcal K} = 1.5$, $s=0.3$, $\phi = 10^{-5}$, $\eta = 0.025$, $\gamma = 0.015$, $a =1$, $b=5$, $c=0.05$, $d=0.25$, $\alpha = 10^{4}$. 
The length of the 'sample' was $N=3000$ points (and $L/N=1$). 
Periodic boundary condition where enforced.
The simulation time was 30000, steps.
Very similar results are obtained in a very broad range of the parameters. 
The one above where choose on the basis of aesthetic considerations only.

%

\begin{thebibliography}{99}
\bibitem{typical}
Although this is the typical behavior for $Im(\omega({\mathbf k}))$, some other more complex behaviors can be observed for special ranges of the parametes.
\end{thebibliography}
%



\end{document}